\documentclass[onecollarge,natbib]{svjour2}
\bibpunct{[}{]}{;}{n}{}{,} 
\smartqed  
\usepackage{graphicx,amsmath,amssymb}
\usepackage[usenames]{color}
\usepackage[colorlinks=true, urlcolor=navyblue, linkcolor=navyblue, citecolor=navyblue]{hyperref}

\definecolor{navyblue}{rgb}{0,0.08,0.45}
\newcommand{\mbf}[1]{\mathbf{#1}}
\newcommand{\half}{{\frac{1}{2}}}
\renewcommand{\bar}[1]{\overline{#1}}

\journalname{Few-Body Systems}
\begin{document}

\title{Light-Front Holography, Light-Front Wavefunctions, and Novel QCD Phenomena}
\thanks{Invited Talks, Presented at LC2011 at Southern Methodist University,  to be published in {\it Few-Body Physics}.}

\titlerunning{Light-Front Holography and QCD }        

\author{Stanley J. Brodsky     \and
        Guy F. de Teramond }

\institute{Stanley J. Brodsky  \at
              Stanford National Accelerator Laboratory \\
                Stanford University, Stanford, CA 94309, USA, and \\
CP$^3$-Origins,
Southern Denmark University, Odense, Denmark\\
              \email{sjbth@slac.stanford.edu}           
           \and
            Guy de Teramond \at Universidad de Costa Rica, San Jos\'e, Costa Rica\\
           \email{gdt@asterix.crnet.cr}          
              }
\date{Received: date / Accepted: date}

\maketitle

\begin{abstract}

Light-Front Holography  is one of the most remarkable features of the AdS/CFT correspondence.
 In spite of its present limitations it provides important physical insights into the non-perturbative regime of QCD and its transition to the perturbative domain.
This novel framework  allows hadronic amplitudes
in a higher dimensional anti-de Sitter (AdS) space 
 to be mapped to frame-independent light-front wavefunctions of hadrons in physical space-time. The model leads to an effective  confining light-front  QCD Hamiltonian and a single-variable light-front Schr\"odinger equation which determines the eigenspectrum and the light-front wavefunctions of hadrons for general spin and orbital angular momentum. The coordinate $z$ in AdS space is uniquely identified with  a Lorentz-invariant  coordinate $\zeta$ which measures the separation of the constituents within a hadron at equal light-front time
 and determines the off-shell dynamics of the bound-state wavefunctions, and thus the fall-off as a function of  the invariant mass of the constituents.  
 The soft-wall holographic model modified by a positive-sign dilaton metric, 
leads to a remarkable one-parameter description of nonperturbative hadron dynamics --  a semi-classical frame-independent first approximation to the spectra and light-front wavefunctions of meson and baryons.  The model predicts 
a Regge spectrum of linear trajectories with the same slope in the leading orbital angular momentum $L$ of hadrons and the radial  quantum number $n.$    The hadron eigensolutions projected on the free Fock basis  provides  the complete  set of valence and non-valence  light-front Fock state wavefunctions 
$\Psi_{n/H}(x_i, \mbf{k}_{\perp i}, \lambda_i)$ which describe the hadron's momentum and spin distributions needed to compute  the direct measures of hadron structure at the quark and gluon level, such as elastic and transition form factors, distribution amplitudes, structure functions, generalized parton distributions and transverse momentum distributions. The effective confining potential also creates quark- antiquark pairs from the amplitude $q \to q \bar q q$. Thus in holographic QCD higher Fock states can have any number of extra $q \bar q$ pairs.
We discuss the relevance of higher Fock-states for describing the detailed structure of space and time-like form factors.
The AdS/QCD model can be systematically improved  by using its complete orthonormal solutions to diagonalize the full QCD light-front Hamiltonian or by applying the Lippmann-Schwinger method in order to systematically include the QCD interaction terms.  A new perspective on quark and gluon condensates is also obtained.

\end{abstract}

\thanks{Presented by SJB and GdT at LIGHTCONE 2011, 23 - 27 May, 2011, Dallas.}


\section{Introduction}
\label{intro}
A number of experimental observables have been identified as  key measures of hadron structure at the quark and gluon level.  In addition to traditional measures such as structure functions, form factors, and distribution amplitudes,  these  include the generalized parton distributions (GPDs) which are measureable in deeply virtual Compton scattering (DVCS)  $\gamma^* p \to \gamma p$,  transverse moment distributions (TMDs), Wigner distributions,  as well  as the $T$-odd spin correlations and transversity distributions such as the Sivers, Collins, and the Boer-Mulders effects.  One can interpret these Lorentz-frame-independent observables in terms of three-dimensional quark and gluon spin and momentum distributions of the hadrons  at fixed light-front time $\tau = t + z/c$. 

The central question is how one can compute such hadronic measures from first principles; i.e., directly from the QCD Lagrangian. Lattice gauge theory can provide some answers, but only indirectly through moments of distributions which can be derived in Euclidean space. Gluonic degrees of freedom are represented as lattice links rather than field theoretic quanta.  The covariant Bethe-Salpeter and Dyson-Schwinger methods have led to many important insights, such as the infrared fixed point behavior of the strong coupling constant and
 ``in-hadron" condensates; however, in practice,  the analyses are limited to ladder approximation in Landau gauge. 

In the case of atomic physics, one computes the distributions within an atom from its Schr\"odinger wavefunction, the eigenfunction of the nonrelativistic approximation  to the exact QED Hamiltonian.  In principle, one could  calculate hadronic spectroscopy and wavefunctions by solving for the eigenstates of the QCD Hamiltonian: 
$H \vert  \Psi \rangle = E \vert \Psi \rangle$
 at fixed time $t.$ However, this traditional method -- called the ``instant" form" by Dirac,~\cite{Dirac:1949cp} is  plagued by complex vacuum and relativistic effects, as well 
as  by  the fact that the boost of such fixed-$t$ wavefunctions away from the hadron's rest frame is an intractable dynamical problem.  A further difficulty~\cite{Hoodbhoy:1998bt}
 is that the concept of a gluon is gauge-dependent -- in general one must include a gauge link (Wilson line) attached to every quark and gluon.

In fact, there is an extraordinarily powerful non-perturbative alternative -- quantization at fixed light-front (LF) time $\tau = t + z/c = x^+ = x^0 + x^3$ -- the ``front-form" of Dirac.~\cite{Dirac:1949cp} In this framework each hadron $H$ is identified as an eigenstate of the QCD Hamiltonian 
$H_{LF}^{QCD} \vert \Psi_H \rangle = M^2_H \vert \Psi_H \rangle$.   Here $H_{LF}^{QCD} = P_\mu P^\mu= P^- P^+ - \mbf{P}^2_\perp$ is derived directly from the QCD Lagrangian or action. The eigenvalues of this Heisenberg equation give the complete mass spectrum of hadrons. The eigensolution  $|\Psi_H \rangle$  projected on the free Fock basis  provides  the complete  set of valence and non-valence  light-front Fock state wavefunctions $\Psi_{n/H}(x_i, \mbf{k}_{\perp i}, \lambda_i)$, which describe the hadron's momentum and spin distributions needed to compute all of the direct measures of hadron structure at the quark and gluon level.

The hadron observables, the structure functions, form factors, distribution amplitudes,  GPDs, TMDs, and Wigner distributions can be computed as convolutions of light-front wavefunctions (LFWFs).  In the case of the GPDs at nonzero skewness, one must include overlaps of LFWFs with different number of Fock constituents.
The $T$-odd spin observables such as the Sivers, Collins, and the Boer-Mulders effects, as well as the distributions  measured in diffractive deep inelastic lepton scattering $\gamma^* p \to p X$,  require the inclusion of initial or final rescattering effects analogous to the Coulomb phase in QED.

The LF Hamiltonian $H_{LF}^{QCD}$ and its LFWF eigensolutions are frame-independent -- no boosts are needed. Furthermore, 
if one quantizes in light-cone gauge $A^+= A^0 + A^3=0$, the gluons have physical polarization $S^z = \pm 1$, there are no ghosts, and the Wilson line reduces to unity, so that  one has a physical interpretation of quark and gluon constituents  and their properties. The empirical observation that quarks carry only a small fraction of the nucleon angular momentum highlights the importance of quark orbital angular momentum.  In fact the nucleon anomalous moment and the Pauli form factor are zero unless the quarks carry nonzero $L^z$.

A remarkable feature of LFWFs is the fact that they are frame
independent; i.e., the form of the LFWF is independent of the
hadron's total momentum $P^+ = P^0 + P^3$ and $\mbf{P}_\perp.$
The simplicity of Lorentz boosts of LFWFs contrasts dramatically with the complexity of the boost of wavefunctions defined at fixed time $t.$~\cite{Brodsky:1968ea}  
Light-front quantization is thus the ideal framework to describe the
structure of hadrons in terms of their quark and gluon degrees of freedom.  The
constituent spin and orbital angular momentum properties of the
hadrons are also encoded in the LFWFs.  
The total  angular momentum projection~\cite{Brodsky:2000ii} 
$J^z = \sum_{i=1}^n  S^z_i + \sum_{i=1}^{n-1} L^z_i$ 
is conserved Fock-state by Fock-state and by every interaction in the LF Hamiltonian.
The constituent spin and orbital angular momentum properties of the hadrons are thus encoded in their LFWFs.

In fact, one can solve the LF Hamiltonian problem for QCD in one-space and one-time  by Heisenberg matrix diagonalization. The complete set of discrete and continuum eigensolutions of mesons and baryons  in QCD(1+1) can be obtained to any desired precision for general color,  multiple flavors, and general quark masses using the discretized light-cone quantized (DLCQ) method.~\cite{Pauli:1985ps,Hornbostel:1988fb}  The  DLCQ approach could in principle be applied to QCD(3+1); however,  in practice, the  huge matrix diagonalization problem is computational challenging, but it may be possible in the future using exascale computer technology.

There is in fact another nonperturbative QCD  approach which leads to an elegant analytical
 and phenomenologically compelling approximation to the full LF Hamiltonian method-- ``Light-Front Holography".~\cite{deTeramond:2008ht}
 Light-Front Holography  is in fact one of the most remarkable features of 
 the AdS/CFT correspondence.~\cite{Maldacena:1997re}
 A precise gravity dual to QCD is not known, but the mechanisms of
confinement can be incorporated into the gauge/gravity
correspondence by modifying the anti-de Sitter  (AdS) geometry in the  large
infrared  domain  $z \sim \frac{1}{\Lambda_{\rm QCD}}$, where 
$\Lambda_{\rm QCD} $ is the scale of the strong
interactions.~\cite{Polchinski:2001tt}  
The resulting dual theory incorporates an
ultraviolet conformal limit at the AdS boundary at $z \to 0$, as
well as the modifications of the background geometry in the large $z$
infrared region  which incorporate confinement. Despite its limitations, this approach to  the gauge/gravity duality, called AdS/QCD, has provided important physical  insights into the non-perturbative dynamics of hadrons -- in some cases, it is the only tool available.

Light-Front Holography provides a rigorous
connection between the description of hadronic modes in 
a higher dimensional AdS space and
the Hamiltonian formulation of QCD in physical space-time quantized
on the light-front at equal light-front time  $\tau$.~\cite{deTeramond:2008ht}  
The physical  states in AdS space are
represented by normalizable modes $\Phi_P(x, z) = e^{-iP \cdot x} \Phi(z)$,
with plane waves along Minkowski coordinates $x^\mu$ and a profile function $\Phi(z)$ 
along the holographic coordinate $z$. The hadronic invariant mass 
$P_\mu P^\mu = M^2$  is found by solving the eigenvalue problem for the
AdS wave equation. Each  light-front hadronic state $\vert \psi(P) \rangle$ is dual to a normalizable string mode $\Phi_P(x,z)$. 
For fields near the AdS boundary the behavior of $\Phi(z)$
depends on the scaling dimension of the corresponding interpolating operators.  Thus each hadron is identified by the twist of its interpolating operator at $z \to 0.$

We have shown that there exists a precise correspondence between the matrix elements of the electromagnetic current and the energy-momentum tensor of the fundamental hadronic constituents in QCD, with the corresponding transition amplitudes describing the interaction of string modes in AdS space with the external sources which propagate in the AdS interior.  The agreement of the results for both 
electromagnetic~\cite{Brodsky:2006uqa, Brodsky:2007hb} and gravitational~\cite{Brodsky:2008pf} hadronic transition amplitudes provides an important consistency test and verification of holographic mapping from AdS to physical observables defined on the light-front.

Light front holographic methods
allows one to project the functional dependence of the wavefunction $\Phi(z)$ computed  in the  AdS fifth dimension to the  hadronic frame-independent light-front wavefunction $\psi(x_i, \mbf{b}_{\perp i})$ in $3+1$ physical space-time. The variable $z $ maps  to a transverse
LF variable $ \zeta(x_i, \mbf{b}_{\perp i})$. 
The result is a single-variable light-front Schr\"odinger equation which determines the eigenspectrum and the LFWFs of hadrons for general spin and orbital angular momentum.  The transverse coordinate $\zeta$ is closely related to the invariant mass squared  of the constituents in the LFWF  and its off-shellness  in  the LF kinetic energy,  and it is thus the natural variable to characterize the hadronic wavefunction.  In fact $\zeta$ is the only variable to appear 
in the relativistic light-front Schr\"odinger equations predicted from 
holographic QCD  in the limit of zero quark masses. 
The coordinate $z$ in AdS space is thus uniquely identified with  a Lorentz-invariant  coordinate $\zeta$ which measures the separation of the constituents within a hadron at equal light-front time.
 This correspondence  agrees
with the intuition that $z$ is related inversely to the internal relative momentum, but the
relation $z \to \zeta$ is precise and exact.

The hadron eigenstates generally have components with different orbital angular momentum; e.g.,  the proton eigenstate in LF holographic QCD  with massless quarks has $L=0$ and $L=1$ light-front Fock components with equal probability.   Higher Fock states with extra quark-anti quark pairs also arise.   The resulting LFWFs then lead to a new range of hadron phenomenology,  including the possibility to compute the hadronization of quark and gluon jets at the amplitude level. The soft-wall model also predicts the form of the non-perturbative effective coupling and its $\beta$-function.~\cite{Brodsky:2010ur}  
 
To a first
semiclassical approximation, where quantum loops and quark masses
are not included, LF holography leads to a LF Hamiltonian equation which
describes the bound-state dynamics of light hadrons  in terms of
the invariant impact variable $\zeta$ at equal light-front
time $\tau = x^0 + x^3$~\cite{Dirac:1949cp, deTeramond:2008ht} 
Remarkably, the AdS equations correspond to the kinetic energy terms of the partons inside a hadron,
whereas the interaction terms,  derived from the modified AdS space, builds confinement and
correspond to the truncation of AdS space in an effective dual
gravity  approximation.~\cite{deTeramond:2008ht}
The effective confining potential also creates quark- antiquark pairs from the amplitude $q \to q \bar q q$. Thus in holographic QCD higher Fock states can have additional $q \bar q$ pairs.
One also obtains a connection between the mass parameter $\mu R$ of the AdS theory with the orbital angular momentum of the constituents in the bound-state eigenstates of $H^{QCD} _{LF}$.
The identification of orbital angular momentum of the constituents is a key element in our description of the internal structure of hadrons using holographic principles, 
since hadrons with the same quark content, but different orbital angular momenta, have different masses.

\section{The Physics of Light Front Wavefunctions}
\label{LFWF}

One of the most important theoretical tools in atomic physics is the
Schr\"odinger wavefunction, which describes the quantum-mechanical
structure of  an atomic system at the amplitude level.  Light-front
wavefunctions  play a similar role in quantum chromodynamics, 
providing a fundamental description of the structure and
internal dynamics of hadrons in terms of their constituent quarks
and gluons.  
The LFWFs  $\psi_H(x_i, \mbf{k}_{\perp i}, \lambda_i)$ are the projections of the hadronic eigensolutions 
$\vert  \psi(P) \rangle$ on the eigenstates of the free Hamiltonian,
where $x_i = k^+_i /P^+ $, $\sum_{i=1}^n k^+_i = P^+,$ $\sum_{i=1}^n \mbf{k}_{\perp i} =0$.
The state with the minimum number of constituents is referred to as the valence Fock state.
The LFWFs of bound states in QCD are thus
relativistic generalizations of the Schr\"odinger wavefunctions of
atomic physics, but they are determined at fixed light-front time
 $\tau = x^0 + x^3$.

In the LF formalism all quanta have positive $k^+$. Since the plus
momentum is conserved,  no vacuum processes appear, and thus the vacuum state is trivial and identical to the free vacuum.  
In effect, the LF vacuum is normal-ordered.
The simple structure of the LF vacuum allows an unambiguous
definition of the partonic content of a hadron in QCD and of hadronic light-front wavefunctions, 
which relate its quark
and gluon degrees of freedom to their asymptotic hadronic state. 

Phenomena usually associated with chiral and gluonic vacuum condensates are properties of the higher LF Fock states,~\cite{Casher:1974xd,Brodsky:2009zd}.  In the case of the Higgs model, the effect of the usual Higgs vacuum expectation value is replaced by a constant $k^+=0$ zero mode field.~\cite{Srivastava:2002mw}

The squares of the LFWFs summed over all Fock states define the generalized parton distributions of the hadrons. The structure functions measured in deep inelastic scattering  which satisfy  DGLAP evolution are derived from the squares of the LFWFs integrated over all variables but the struck quark's light-front momentum fraction $x = x_{bj}$.  The overlap of LFWFs defines the GPDs measured in DVCS. 

The gauge-invariant distribution amplitude $\phi(x,Q^2)$ which controls exclusive processes at high $Q^2$  is the integral of the valence LFWF in LCG integrated over the internal transverse momentum $k^2_\perp < Q^2$, because the Wilson line is trivial in this gauge.   The logarithmic evolution of the distribution amplitude satisfies the ERBL evolution equation. It is also possible to quantize QCD in  Feynman gauge in the light front.~\cite{Srivastava:1999gi}

Other advantages of light-front quantization include:

\begin{itemize}

\item
LF Hamiltonian perturbation theory provides a simple method for deriving analytic forms for the analog of Parke-Taylor amplitudes~\cite{Motyka:2009gi} where each particle spin $S^z$ is quantized in the LF $z$ direction.  The gluonic $g^6$ amplitude  $T(-1 -1 \to +1 +1 +1 +1 +1 +1)$  requires $\Delta L^z =8;$ it thus must vanish at tree level since each three-gluon vertex has  $\Delta L^z = \pm 1.$ However, the order $g^8$ one-loop amplitude can be nonzero.

\item
Amplitudes in light-front perturbation theory can be renormalized using the ``alternate denominator'' subtraction method.~\cite{Brodsky:1973kb}  The application to QED has been checked at one and two loops.~\cite{Brodsky:1973kb}

\item 
One can easily show using LF quantization that the anomalous gravitomagnetic moment $B(0)$  of a nucleon, as  defined from the spin flip matrix element of the energy-momentum tensor, vanishes Fock-state by Fock state,~\cite{Brodsky:2000ii} as required by the equivalence principle.~\cite{Teryaev:1999su}

\item
LFWFs obey the cluster decomposition theorem, providing the only proof of this theorem for relativistic bound states.~\cite{Brodsky:1985gs}

\item 
LF quantization provides a distinction between static  (square of LFWFs) distributions versus non-universal dynamic structure functions,  such as the Sivers single-spin correlation and diffractive deep inelastic scattering which involve final state interactions.  The origin of nuclear shadowing and process independent anti-shadowing also becomes explicit.   

\item 
LF quantization provides a simple method to implement jet hadronization at the amplitude level.  

\item 
The instantaneous fermion interaction in LF  quantization provides a simple derivation of the $J=0$
fixed pole contribution to deeply virtual Compton scattering.~\cite{Brodsky:2009bp}

\end{itemize}

\section{Light-Front Bound-State Hamiltonian Equation of Motion and Light-Front Holography \label{LF Hamiltonian}
}

A key step in the analysis of an atomic system such as positronium
is the introduction of the spherical coordinates $r, \theta, \phi$
which  separates the dynamics of Coulomb binding from the
kinematical effects of the quantized orbital angular momentum $L$.
The essential dynamics of the atom is specified by the radial
Schr\"odinger equation whose eigensolutions $\psi_{n,L}(r)$
determine the bound-state wavefunction and eigenspectrum. In our recent 
work, we have shown that there is an analogous invariant
light-front coordinate $\zeta$ which allows one to separate the
essential dynamics of quark and gluon binding from the kinematical
physics of constituent spin and internal orbital angular momentum.
The result is a single-variable LF Schr\"odinger equation for QCD
which determines the eigenspectrum and the light-front wavefunctions
of hadrons for general spin and orbital angular momentum.~\cite{deTeramond:2008ht}  If one further chooses  the constituent rest frame (CRF)~\cite{Danielewicz:1978mk,Karmanov:1979if,Glazek:1983ba}  where $\sum^n_{i=1} \mbf{k}_i \! = \! 0$, then the kinetic energy in the LFWF displays the usual 3-dimensional rotational invariance. Note that if the binding energy is nonzero, $P^z \ne 0,$ in this frame.

One can also derive light-front holography using a first semiclassical approximation  to transform the fixed 
light-front time bound-state Hamiltonian equation of motion in QCD
\begin{equation} \label{LFH}
H_{LF}^{QCD} \vert  \psi(P) \rangle =  M_{H}^2 \vert  \psi(P) \rangle,
\end{equation}
with  $H_{LF}^{QCD} \equiv P_\mu P^\mu  =  P^- P^+ -  \mbf{P}_\perp^2$,
to  a corresponding wave equation in AdS 
space.~\cite{deTeramond:2008ht} To this end we
expand the initial and final hadronic states in terms of its Fock components. We use the 
frame $P = \big(P^+, M^2/P^+, \vec{0}_\perp \big)$ where $H_{LF} =  P^+ P^-$.
We find 
\begin{equation}   
 M_H^2  =  \sum_n  \prod_{j=1}^{n-1} \int d x_j \, d^2 \mbf{b}_{\perp j} \,
\psi_{n/H}^*(x_j, \mbf{b}_{\perp j}) 
  \sum_q   \left(\frac{ \mbf{- \nabla}_{ \mbf{b}_{\perp q}}^2  \! + m_q^2 }{x_q} \right) 
 \psi_{n/H}(x_j, \mbf{b}_{\perp j}) 
  + {\rm (interactions)} , \label{eq:Mba}
 \end{equation}
plus similar terms for antiquarks and gluons ($m_g = 0)$.

 Each constituent of the light-wavefunction  $\psi_{n/H}(x_i, \mbf{k}_{\perp i}, \lambda_i)$  of a hadron is on its respective mass shell 
 $k^2_i= k^+_i k^-_i - \mbf{k}^2_\perp = m^2_i$, $i = 1, 2 \cdots n.$   Thus $k^-= 
{{\mbf k}^2_\perp+  m^2_i\over x_i P^+}$.
However,  the light-front wavefunction represents a state which is off the light-front energy shell: $P^-  - \sum_i^n k^-_n < 0$, for a stable hadron.  Scaling out $P^+ = \sum^n_i k^+_i$, the 
off-shellness of the $n$-parton LFWF is thus $M^2_H -  M^2_n$, where
the invariant mass  of the constituents $M_n$ is
\begin{equation}
 M_n^2  = \Big( \sum_{i=1}^n k_i^\mu\Big)^2 = \sum_i \frac{\mbf{k}_{\perp i}^2 +  m_i^2}{x_i}.
 \end{equation}

The action principle selects the configuration which minimizes the time-integral of the Lagrangian $L= T-V$, thus minimizing the kinetic energy $T$  and maximizing the attractive forces of the potential $V$. Thus in a fixed potential, the light-front wavefunction  peaks at the minimum value of the invariant mass of the constituents; i.e.,  at  the minimum off-shellness $M^2_H -  M^2_n$.   In the case of massive constituents,  the minimum LF off-shellness occurs when all of the constituents have equal rapidity: $x_i \simeq {m^2_{\perp i}\over \sum^n_j m^2_{\perp j}}, $ where $m_{\perp i} = \sqrt {k^2_{\perp i} + m^2_i}.$ This is the central principle underlying the intrinsic heavy sea-quark distributions of hadrons.  The functional dependence  for a given Fock state is given in terms of the invariant mass,  the measure of the off-energy shell of the bound state.  

If we want to simplify further the description of the multiple parton system and reduce its dynamics to a single variable problem, we must take the limit of quark masses to zero. 
Indeed, the underlying classical QCD Lagrangian with massless quarks is scale and conformal invariant,~\cite{Parisi:1972zy} and consequently only in this limit it is possible to map the equations of motion and transition matrix elements to their correspondent conformal AdS  expressions.

 To simplify the discussion we will consider a two-parton hadronic bound state.  In the limit of zero quark masses
$m_q \to 0$
\begin{equation}  \label{eq:Mb}
M^2  =  \int_0^1 \! \frac{d x}{x(1-x)} \int  \! d^2 \mbf{b}_\perp  \,
  \psi^*(x, \mbf{b}_\perp) 
  \left( - \mbf{\nabla}_{ {\mbf{b}}_{\perp}}^2\right)
  \psi(x, \mbf{b}_\perp) +   {\rm (interactions)}.
 \end{equation}
For $n=2$, $M_{n=2}^2 = \frac{\mbf{k}_\perp^2}{x(1-x)}$. 
Similarly in impact space the relevant variable for a two-parton state is  $\zeta^2= x(1-x)\mbf{b}_\perp^2$.
Thus, to first approximation  LF dynamics  depend only on the boost invariant variable
$M_n$ or $\zeta,$
and hadronic properties are encoded in the hadronic mode $\phi(\zeta)$ from the relation
\begin{equation} \label{eq:psiphi}
\psi(x,\zeta, \varphi) = e^{i M \varphi} X(x) \frac{\phi(\zeta)}{\sqrt{2 \pi \zeta}} ,
\end{equation}
thus factoring out the angular dependence $\varphi$ and the longitudinal, $X(x)$, and transverse mode $\phi(\zeta)$
with normalization $ \langle\phi\vert\phi\rangle = \int \! d \zeta \,
 \vert \langle \zeta \vert \phi\rangle\vert^2 = 1$.
 The factorization of the LFWF given by (\ref{eq:psiphi}) is a natural factorization in the light front formalism since the
corresponding canonical generators, the longitudinal and transverse generators $P^+$ and $\mbf{P}_\perp$ and the $z$-component of the orbital angular momentum
$J^z$, are kinematical generators which commute with the LF Hamiltonian generator $P^-$.~\cite{Dirac:1949cp}
 The mapping  of transition matrix elements
 for arbitrary values of the momentum transfer described in Sec. \ref{EMFF} gives $X(x) = \sqrt{x(1-x)}$~\cite{Brodsky:2006uqa,Brodsky:2007hb,Brodsky:2008pf} in the limit of zero quark masses.

We can write the Laplacian operator in (\ref{eq:Mb}) in circular cylindrical coordinates $(\zeta, \varphi)$
and factor out the angular dependence of the
modes in terms of the $SO(2)$ Casimir representation $L^2$ of orbital angular momentum in the
transverse plane. Using  (\ref{eq:psiphi}) we find~\cite{deTeramond:2008ht}
\begin{equation} \label{eq:KV}
M^2   =  \int \! d\zeta \, \phi^*(\zeta) \sqrt{\zeta}
\left( -\frac{d^2}{d\zeta^2} -\frac{1}{\zeta} \frac{d}{d\zeta}
+ \frac{L^2}{\zeta^2}\right)
\frac{\phi(\zeta)}{\sqrt{\zeta}}   \\
+ \int \! d\zeta \, \phi^*(\zeta) \, U(\zeta)  \, \phi(\zeta) ,
\end{equation}
where all the complexity of the interaction terms in the QCD Lagrangian is summed up in the effective potential $U(\zeta)$.
The light-front eigenvalue equation $H_{LF} \vert \phi \rangle = M^2 \vert \phi \rangle$
is thus a LF wave equation for $\phi$
\begin{equation} \label{LFWE}
\left(-\frac{d^2}{d\zeta^2}
- \frac{1 - 4L^2}{4\zeta^2} + U(\zeta) \right) 
\phi(\zeta) = M^2 \phi(\zeta),
\end{equation}
an effective single-variable light-front Schr\"odinger equation which is
relativistic, covariant and analytically tractable. Using (\ref{eq:Mb}) one can readily
generalize the equations to allow for the kinetic energy of massive
quarks.~\cite{Brodsky:2008pg}  In this case, however,
the longitudinal mode $X(x)$ does not decouple from the effective LF bound-state equations.

We now compare (\ref{LFWE}) with the wave equation in AdS$_{d+1}$ in presence of a dilaton background $\varphi(z)$ for a spin-$J$ mode $\Phi_J$, $\Phi_J= \Phi_{\mu_1 \mu_2 \cdots \mu_J}$, with all the polarization indices
along the physical 3 + 1 coordinates~\cite{deTeramond:2008ht, deTeramond:2010ge}
\footnote{A detailed discussion of higher integer and half-integer spin wave equations  in modified AdS spaces
will be given in~\cite{BDdT:2011}. See also the discussion in Ref. \cite{Gutsche:2011vb}.}

\begin{equation} \label{WeJ}
\left[-\frac{ z^{d-1 -2 J}}{e^{\varphi(z)}}   \partial_z \left(\frac{e^{\varphi(z)}}{z^{d-1 - 2 J}} \partial_z\right)
+ \left(\frac{\mu R}{z}\right)^2\right] \Phi_{\mu_1 \mu_2 \cdots \mu_J} = M^2 \Phi_{\mu_1 \mu_2 \cdots \mu_J}.
\end{equation}

Upon the substitution $z \! \to\! \zeta$  and
$\phi_J(\zeta)   = \left(\zeta/R\right)^{-3/2 + J} e^{\varphi(z)/2} \, \Phi_J(\zeta)$,
in (\ref{WeJ}), we find for $d=4$ the QCD light-front wave equation (\ref{LFWE}) with effective potential~\cite{deTeramond:2010ge}
\begin{equation} \label{U}
U(\zeta) = \half \varphi''(z) +\frac{1}{4} \varphi'(z)^2  + \frac{2J - 3}{2 z} \varphi'(z) ,
\end{equation}
where the fifth dimensional mass $\mu$ is not a free parameter but scales as $(\mu R)^2 = - (2-J)^2 + L^2$. If $L^2 \ge 0$ the LF Hamiltonian is positive definite
 $\langle \phi \vert H_{LF} \vert \phi \rangle \ge 0$ and thus $ M^2 \ge 0$.
 If $L^2 < 0$ the bound state equation is unbounded from below. The critical value corresponds to $L=0$.
 The quantum mechanical stability $L^2 >0$ for $J=0$ is thus equivalent to the
 Breitenlohner-Freedman stability bound in AdS.~\cite{Breitenlohner:1982jf}

In the hard-wall model one has $U(z)=0$; confinement is introduced by requiring the wavefunction to vanish at $z=z_0 \equiv 1/\Lambda_{\rm QCD}.$~\cite{Polchinski:2001tt}
In the case of the soft-wall model,~\cite{Karch:2006pv}  the potential arises from a  ``dilaton'' modification of the AdS metric; it  has the form of a harmonic oscillator. For the confining  positive-sign dilaton background $\exp(+ \kappa^2 z^2)$~\cite {deTeramond:2009xk, Andreev:2006ct}  we find the effective potential 
$U(z) = \kappa^4 z^2 + 2 \kappa^2(L+S-1)$ where $J=L+S$. The resulting mass spectra  for mesons  at zero quark mass is
${\cal M}^2 = 4 \kappa^2 (n + L +S/2)$.

\begin{figure}[h]
\begin{center}
\includegraphics[width=7.2cm]{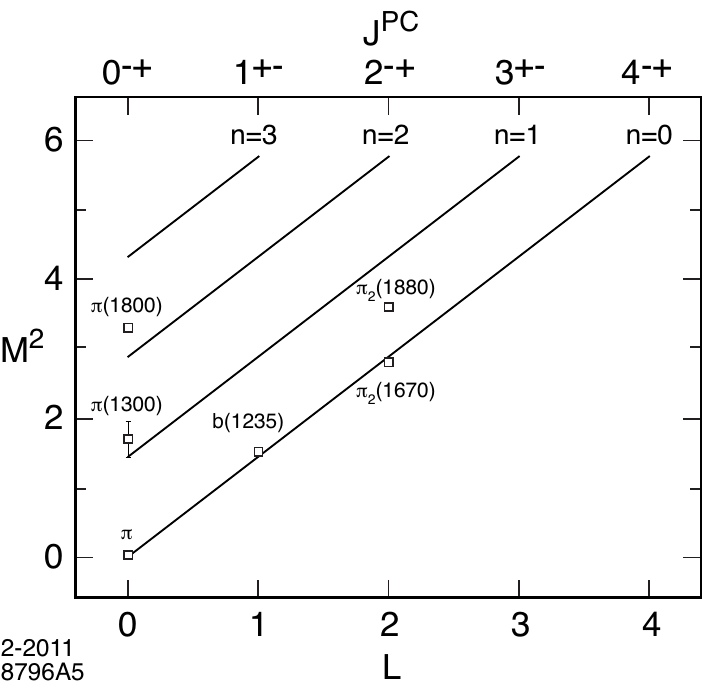}  \hspace{10pt}
\includegraphics[width=7.2cm]{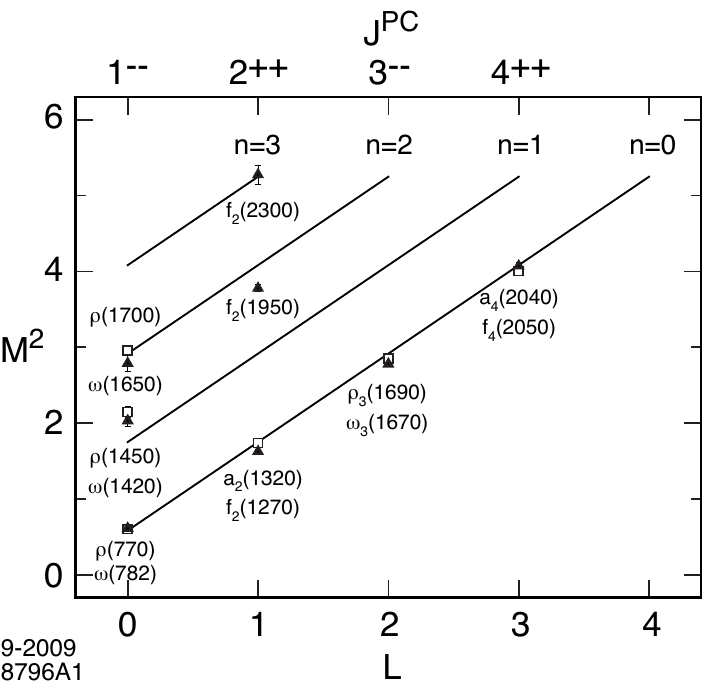}
 \caption{Parent and daughter Regge trajectories for (a) the $S=0$ $\pi$-meson family with
$\kappa= 0.6$ GeV; and (b) the   $S=1$, $I\!=\!1$, $\rho$-meson
 and $S=1$, $I\!=\!0$,  $\omega$-meson families with $\kappa= 0.54$ GeV. Only confirmed PDG states~\cite{Amsler:2008xx} are shown.}
\label{pionspec}
\end{center}
\end{figure}

The spectral predictions for  light meson and vector meson  states are compared with experimental data
in Fig. \ref{pionspec} for the positive sign dilaton model discussed here.

\section{Light-Front Holography and Form Factors
\label{EMFF}}

A form factor in QCD is defined by the transition matrix element of a local quark current between hadronic states.  In AdS space form factors are computed from the overlap integral of normalizable modes with boundary currents which propagate in AdS space. The AdS/CFT duality incorporates the connection between the twist scaling dimension of the  QCD boundary interpolating operators
to the falloff 
of the 
normalizable modes in AdS near its conformal boundary. If both quantities represent the same physical observable for any value of 
the transferred momentum squared $q^2$, 
a precise correspondence can be established between the string modes $\Phi$ in AdS space and the light front wavefunctions of hadrons $\psi_{n/H}$ in physical four dimensional space-time.~\cite{Brodsky:2006uqa} 
The same results follow from comparing the relativistic light-front Hamiltonian equation describing bound states in QCD with the wave equations describing the propagation of modes in a warped AdS 
space as shown in the previous section.~\cite{deTeramond:2008ht}

Light-Front Holography can be derived by observing the correspondence between matrix elements obtained in AdS/CFT with the corresponding formula using the light-front
representation.~\cite{Brodsky:2006uqa}
In the higher dimensional gravity theory, the hadronic matrix element  corresponds to
the  non-local coupling of an external electromagnetic field $A^M(x,z)$  propagating in AdS with the extended mode $\Phi(x,z)$~\cite{Polchinski:2002jw}
 \begin{equation} \label{FFAdSLF}
 \int d^4x \, dz  \, A^{M}(x,z)
 \Phi^*_{P'}(x,z) \overleftrightarrow\partial_M \Phi_P(x,z)  \sim
 (2 \pi)^4 \delta^4 \left( P'  \! - P - q\right) \epsilon_\mu  \langle \psi(P')  \vert J^\mu \vert \psi(P) \rangle ,
 \end{equation}
 where the coordinates of AdS$_5$ are the Minkowski coordinates $x^\mu$ and $z$ labeled $x^M = (x^\mu, z)$, with
$M = 1, \cdots 5$,
 and $g$ is the determinant of the metric tensor. The expression on the right-hand side  represents the QCD EM transition amplitude in physical space-time. It is the EM matrix element of the quark current  $J^\mu = e_q \bar q \gamma^\mu q$, and represents a local coupling to pointlike constituents. Although the expressions for the transition amplitudes look very different, one can show  that a precise mapping of the $J^+$ elements  can be carried out at fixed light-front time.

The light-front electromagnetic form factor in impact
space~\cite{Brodsky:2006uqa,Brodsky:2007hb,Soper:1976jc} can be written as a sum of overlap of light-front wave functions of the $j = 1,2, \cdots, n-1$ spectator
constituents:
\begin{equation} \label{eq:FFb}
F(q^2) =  \sum_n  \prod_{j=1}^{n-1}\int d x_j d^2 \mbf{b}_{\perp j}   \sum_q e_q
            \exp \! {\Bigl(i \mbf{q}_\perp \! \cdot \sum_{j=1}^{n-1} x_j \mbf{b}_{\perp j}\Bigr)}
 \left\vert  \psi_{n/H}(x_j, \mbf{b}_{\perp j})\right\vert^2 ,
\end{equation}
where the normalization is defined by
\begin{equation}  \label{eq:Normb}
\sum_n  \prod_{j=1}^{n-1} \int d x_j d^2 \mathbf{b}_{\perp j}
\vert \psi_{n/H}(x_j, \mathbf{b}_{\perp j})\vert^2 = 1.
\end{equation}

The formula (\ref{eq:FFb}) is exact if the sum is over all Fock states $n$.~\cite{Drell:1969km, West:1970av}
For definiteness we shall consider
the $\pi^+$  valence Fock state
$\vert u \bar d\rangle$ with charges $e_u = \frac{2}{3}$ and $e_{\bar d} = \frac{1}{3}$.
For $n=2$, there are two terms which contribute to the $q$-sum in (\ref{eq:FFb}).
Exchanging $x \leftrightarrow 1 \! - \! x$ in the second integral  we find
\begin{equation}  \label{eq:PiFFb}
 F_{\pi^+}(q^2)  =  2 \pi \int_0^1 \! \frac{dx}{x(1-x)}  \int \zeta d \zeta \,
J_0 \! \left(\! \zeta q \sqrt{\frac{1-x}{x}}\right)
\left\vert \psi_{u \bar d/ \pi}\!(x,\zeta)\right\vert^2,
\end{equation}
where $\zeta^2 =  x(1  -  x) \mathbf{b}_\perp^2$ and $F_{\pi^+}(q\!=\!0)=1$.

We now compare this result with the electromagnetic (EM) form-factor
in  AdS  space time. 
The incoming electromagnetic field propagates in AdS according to
$A_\mu(x^\mu ,z) = \epsilon_\mu(q) e^{-i q \cdot x} V(q^2, z)$,
where $V(q^2,z)$, the bulk-to-boundary propagator, is the solution of the AdS wave equation
with boundary conditions $V(q^2 = 0, z ) = V(q^2, z = 0) = 1$.~\cite{Polchinski:2002jw}
The propagation of the pion in AdS space is described by a normalizable mode
$\Phi_P(x^\mu, z) = e^{-i P  \cdot x} \Phi(z)$ with invariant  mass $P_\mu P^\mu = \mathcal{M}_\pi^2$ and plane waves along Minkowski coordinates $x^\mu$.  
Factoring out the plane wave dependence of the AdS fields  we find
the transition amplitude   $(Q^2 = - q^2 >0$)
\begin{equation}
\langle P' \vert J^\mu \vert P \rangle = \left(P + P' \right)^\mu F(Q^2),
\end{equation}
where we have extracted the overall factor  $ (2 \pi)^4 \delta^4 \left( P'  \! - P - q\right)$ from momentum
conservation at the vertex from integration over Minkowski variables in (\ref{FFAdSLF}).  We find for $F(Q^2)$~\cite{Polchinski:2002jw}
\begin{equation}
F(Q^2) = R^3 \int \frac{dz}{z^3} \, V(Q^2, z) \vert \Phi(z) \vert^2,
\label{eq:FFAdS}
\end{equation}
where $F(Q^2 = 0) = 1$. 
Using the integral representation of   $V(Q^2, z)$ 
\begin{equation} \label{eq:intJ}
V(Q^2, z) =  z Q K_1(z Q)=  \int_0^1 \! dx \, J_0\negthinspace \left(\negthinspace\zeta Q
\sqrt{\frac{1-x}{x}}\right) ,
\end{equation}
 we write the AdS electromagnetic form-factor as
\begin{equation}
F(Q^2)  =    R^3 \! \int_0^1 \! dx  \! \int \frac{dz}{z^3} \,
J_0\!\left(\!z Q\sqrt{\frac{1-x}{x}}\right) \left \vert\Phi(z) \right\vert^2 .
\label{eq:AdSFx}
\end{equation}
To compare with  the light-front QCD  form factor expression (\ref{eq:PiFFb})  we 
use the factorization of the LFWF given by (\ref{eq:psiphi}).
If both expressions for the form factor are identical for arbitrary values of $Q$,
we obtain $\phi(\zeta) = (\zeta/R)^{3/2} \Phi(\zeta)$ and $X(x) = \sqrt{x(1-x)}$,~\cite{Brodsky:2006uqa}
where we identify the transverse impact LF variable $\zeta$ with the holographic variable $z$,
$z \to \zeta = \sqrt{x(1-x)} \vert \mbf b_\perp \vert$. The normalization of the LFWF mode $\phi(\zeta)$ is
 $ \langle\phi\vert\phi\rangle = \int \! d \zeta \,
 \vert \langle \zeta \vert \phi\rangle\vert^2 = 1$.

Extension of the results to arbitrary $n$ follows from the $x$-weighted definition of the
transverse impact variable of the $n-1$ spectator system:~\cite{Brodsky:2006uqa}
\begin{equation} \label{zeta}
\zeta = \sqrt{\frac{x}{1-x}} ~ \Big\vert \sum_{j=1}^{n-1} x_j \mbf{b}_{\perp j} \Big\vert ,
\end{equation}
where $x = x_n$ is the longitudinal
momentum fraction of the active quark.

Conserved currents are not renormalized and correspond to five dimensional massless fields propagating in AdS according to the relation
$(\mu R)^2 = (\Delta - p) (\Delta + p -  4)$  for a $p$ form in $d=4$. In the usual AdS/QCD framework~\cite{Erlich:2005qh, DaRold:2005zs} this  corresponds to $\Delta = 3$ or 1, the canonical dimensions of
an EM current and the massless gauge field respectively.  Normally one uses a hadronic  interpolating operator  with minimum twist $\tau$ to identify a hadron in AdS/QCD and to predict the power-law fall-off behavior of its form factors and other hard
scattering amplitudes;~\cite{Polchinski:2001tt}  e.g.,  for a two-parton bound state $\tau = 2$.   However, in the case of a current, one needs to  use  an effective field operator  with dimension $\Delta =3.$ The apparent inconsistency between twist and dimension is removed by noticing that in the light-front one chooses to calculate the  matrix element of the twist-3 plus  component of the current  $J^+$,~\cite{Brodsky:2006uqa, Brodsky:2007hb} in order to avoid coupling to Fock states with different numbers of constituents.

\begin{figure}[h]
\includegraphics[angle=0,width=8.0cm]{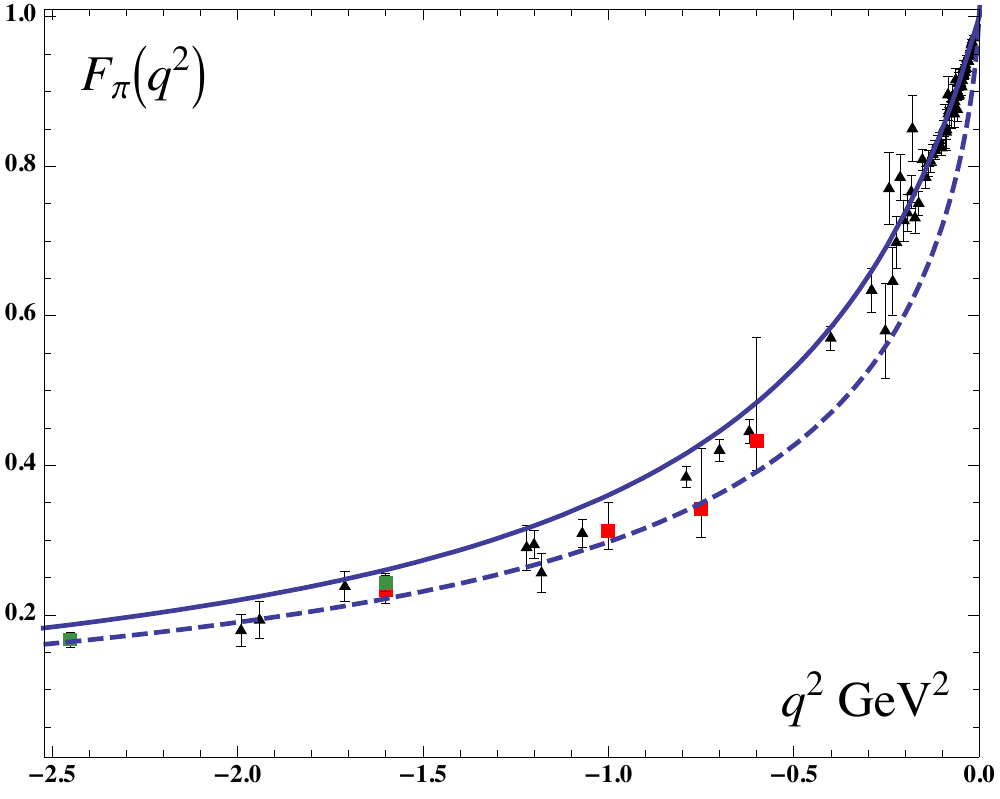}
\caption{ \small Space-like pion form factor $F_\pi(q^2)$.  Continuous line: confined current; dashed  line: free current.
Triangles are the data compilation  from Baldini,~\cite{Baldini:1998qn}  boxes  are JLAB data.~\cite{Tadevosyan:2007yd, Horn:2006tm}}
 \label{PionFFSL}
\end{figure}

The results described above correspond to a ``free" current propagating on AdS space and dual to the EM pointlike current in the DYW light-front formula,  which allow us to map state-by-state.\footnote{In general the mapping relates the AdS density $\Phi^2(z)$ to an  effective LF single particle transverse density.~\cite {Brodsky:2006uqa}}
This mapping has the shortcoming that the pole structure of the form factor is not built on the timelike region. Furthermore, the moments of the form factor at $Q^2=0$ diverge, giving for example an infinite charge radius.  The pole structure is generated when the EM current is confined, this means, when the EM current propagates on
a IR deformed AdS space to mimic confinement. This also leads to finite moments at $Q^2=0$, as illustrated on Fig. \ref{PionFFSL}. The effective ``dressed''  current encodes  non-perturbative dynamical aspects that cannot be learned from a term-by-term holographic mapping, unless one adds an infinite number of terms.

Hadronic form factors  for the harmonic potential $\kappa^2 z^2$  have a simple analytical form~\cite{Brodsky:2007hb} 
\begin{equation} \label{Ftau}   
 F_\tau(Q^2) =  \frac{1}{{\Big(1 + \frac{Q^2}{M^2_\rho} \Big) }
 \Big(1 + \frac{Q^2}{M^2_{\rho'}}  \Big)  \cdots 
       \Big(1  + \frac{Q^2}{M^2_{\rho^{\tau-2}}} \Big)} ,
\end{equation}
which is  expressed as a $\tau - 1$ product of poles along the vector meson Regge radial trajectory.
For a pion, for example, the lowest Fock state -- the valence state -- is a twist-2 state, and thus the form factor is the well known monopole form.~\cite{Brodsky:2007hb}
The remarkable analytical form of (\ref{Ftau}),
expressed in terms of the $\rho$ vector meson mass and its radial excitations, incorporates the correct scaling behavior from the constituent's hard scattering with the photon and the mass gap from confinement.  For a confined EM current in AdS a precise mapping can also be carried out to the DYW expression for the form factor. In this case we we find an effective LFWF, which corresponds to a superposition of an infinite number of Fock states.~\cite{Brodsky:2011xx}

Light front holography provides a precise relation of the fifth-dimensional mass $\mu$ with the total  and orbital angular momentum of a hadron in the transverse LF plane $(\mu R)^2 = - (2 - J)^2 + L^2$, $L = \vert L^z\vert$,~\cite{deTeramond:2008ht} and thus  a conserved EM current  corresponds to poles along the $J=L=1$ radial  trajectory. For the twist-3 computation of the space-like form factor, which involves the current $J^+$, the poles do not correspond to the physical poles of the twist-2 transverse current $\mbf{J}_\perp$ presented
in the annihilation channel, namely the $J=1$, $L=0$ radial trajectory. Consequently, the location of the poles in the final result should be shifted to their physical positions.~\cite{Brodsky:2011xx} When this is done, the results agree extremely well with the space-like pion form factor data as shown in Fig.  \ref{PionFFSL}, as well as for the space-like proton Dirac elastic and transition form factor data as discussed in Sec. \ref{NFF}. 

We have also studied the photon-to-meson transition form factors (TFFs) $F_{M \gamma}(Q^2)$ measured in $\gamma \gamma^* \to M$  reactions using light-front holographic methods,~\cite{Brodsky:2011xx} processes which have been of intense experimental and theoretical interest. 
Holographic QCD methods have also been used to obtain general parton distributions (GPDs) in
 Refs.  \cite{Vega:2010ns} and \cite{Nishio:2011xa}.

\section{Nucleons in Light-Front Holography \label{NucleonSpec}}

For baryons, the light-front wave equation is a linear equation
determined by the LF transformation properties of spin 1/2 states. A linear confining potential
$U(\zeta) \sim \kappa^2 \zeta$ in the LF Dirac
equation leads to linear Regge trajectories.~\cite{Brodsky:2008pg}   For fermionic modes the light-front matrix
Hamiltonian eigenvalue equation $D_{LF} \vert \psi \rangle = M \vert \psi \rangle$, $H_{LF} = D_{LF}^2$,
in a $2 \times 2$ spinor  component
representation is equivalent to the system of coupled linear equations
\begin{eqnarray} \label{eq:LFDirac} \nonumber
- \frac{d}{d\zeta} \psi_- -\frac{\nu+\half}{\zeta}\psi_-
- \kappa^2 \zeta \psi_-&=&
M \psi_+, \\ \label{eq:cD2k}
  \frac{d}{d\zeta} \psi_+ -\frac{\nu+\half}{\zeta}\psi_+
- \kappa^2 \zeta \psi_+ &=&
M \psi_-.
\end{eqnarray}
with eigenfunctions
\begin{eqnarray} \nonumber
\psi_+(\zeta) &\sim& \zeta^{\frac{1}{2} + \nu} e^{-\kappa^2 \zeta^2/2}
  L_n^\nu(\kappa^2 \zeta^2) ,\\  \label{states}
\psi_-(\zeta) &\sim&  \zeta^{\frac{3}{2} + \nu} e^{-\kappa^2 \zeta^2/2}
 L_n^{\nu+1}(\kappa^2 \zeta^2),
\end{eqnarray}
and  eigenvalues
\begin{equation}
M^2 = 4 \kappa^2 (n + \nu + 1) .
\end{equation}

The LF wave equation has also a geometric interpretation: it corresponds to the Dirac equation in AdS$_5$
space in presence of a linear potential $\kappa^2 z$
\begin{equation} \label{eq:DEz}
\left[i\big( z \eta^{M N} \Gamma_M \partial_N + 2 \, \Gamma_z \big) + \kappa^2 z
 + \mu R \right] \Psi = 0 ,
\end{equation}
as can be shown  by using the transformation $\Psi( z) \sim z^2 \psi(z)$, $z \to \zeta$.   The equation of motion 
for baryons in the light front is thus mapped to a Dirac equation for spin-$\half$ modes  in  AdS space.

The baryon interpolating operator
$ \mathcal{O}_{3 + L} =  \psi D_{\{\ell_1} \dots
 D_{\ell_q } \psi D_{\ell_{q+1}} \dots
 D_{\ell_m\}} \psi$,  $L = \sum_{i=1}^m \ell_i$, is a twist 3,  dimension $9/2 + L$ with scaling behavior given by its
 twist-dimension $3 + L$. We thus require $\nu = L+1$ to match the short distance scaling behavior.   One can interpret $L$ as the maximal value of $\vert L^z \vert$ in a LFWF Fock state.
 In the case of massless quarks, the nucleon eigenstates have Fock components with different orbital angular momentum, $L = 0$ and $L = 1$, but  with equal probability.   {\it In effect, the nucleon's angular momentum is carried by quark orbital angular momentum. } Higher spin fermionic modes
 $\Psi _{\mu_1 \cdots \mu_{J-1/2}}$, $J > 1/2$, with all of its polarization indices along the $3+1$ coordinates follow by shifting dimensions for the fields as shown for the case of mesons in Ref.~\cite{deTeramond:2009xk}. 
Thus, as in the meson sector,  the increase  in the 
mass $M^2$ squared for baryonic states for increased radial and orbital quantum numbers is
$\Delta n = 4 \kappa^2$, $\Delta L = 4 \kappa^2$ and $\Delta S = 2 \kappa^2,$ 
relative to the lowest ground state,  the proton; i.e., the slope of the spectroscopic trajectories in $n$ and $L$ are identical .

\begin{figure}[!]
\includegraphics[angle=0,width=14.0cm]{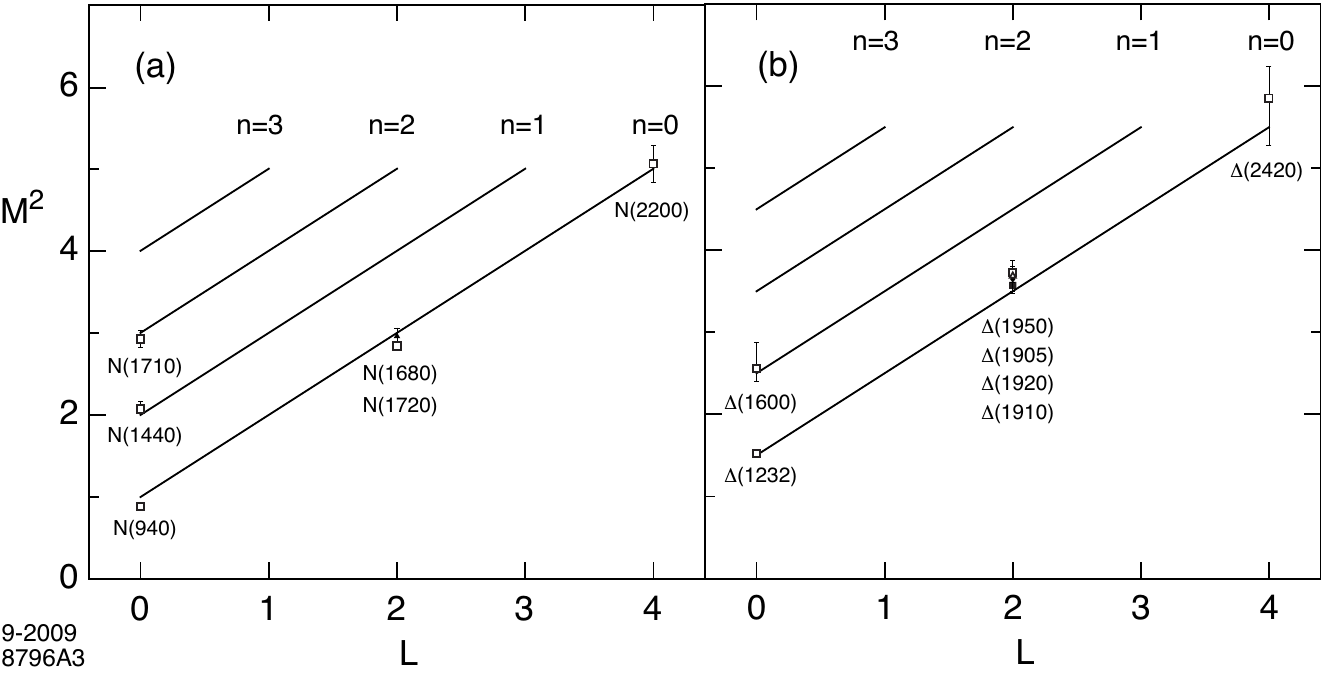}
\caption{Positive parity Regge trajectories for  the  $N$ and $\Delta$ baryon families for $\kappa= 0.5$ GeV}
\label{Baryons}
\end{figure}

We can fix the overall energy scale to be identical for mesons and baryons by imposing chiral symmetry for the pion~\cite{deTeramond:2010we} in the LF Hamiltonian equations: $m^2_\pi =0 $ for $m_q\to 0.$
The resulting predictions for the spectroscopy of positive-parity light baryons  are shown in Fig. \ref{Baryons}.
Only confirmed PDG~\cite{Amsler:2008xx} states are shown.
The Roper state $N(1440)$ and the $N(1710)$ are well accounted for in this model as the first  and second radial
states. Likewise the $\Delta(1660)$ corresponds to the first radial state of the $\Delta$ family. The model is  successful in explaining the parity degeneracy observed in the light baryon spectrum, such as the $L\! =\!2$, $N(1680)\!-\!N(1720)$ degenerate pair and the $L=2$, $\Delta(1905), \Delta(1910), \Delta(1920), \Delta(1950)$ states which are degenerate within error bars.  The parity degeneracy of baryons is also a property of the 
 hard-wall model, but radial states are not well described in this model.~\cite{deTeramond:2005su} 
 
In addition to the Hard Wall model, there are other examples of
modified AdS space. In particular one can solve the five dimensional Einstein-dilaton
equations and obtain linear Regge trajectories and the Wilson Loop
area law for mesons, as is done in ref.~\cite{dePaula:2008fp}
For other calculations of the hadronic spectrum in the framework of AdS/QCD, see Refs.~\cite{Hong:2006ta,  Forkel:2007cm, Nawa:2008xr,  Forkel:2008un, Ahn:2009px,  Zhang:2010bn, Kirchbach:2010dm}.

An important feature of light-front holography is that it predicts the identical multiplicity of states for mesons
and baryons that is observed experimentally.~\cite{Klempt:2007cp} This remarkable property could have a simple explanation in the cluster decomposition of the
holographic variable, which labels a system of partons as an active quark plus a system on $n-1$ spectators. From this perspective, a baryon with $n=3$ looks in light-front holography as a quark--scalar-diquark system.

\section{Nucleon Form Factors \label{NFF}}

In the higher dimensional gravity theory, hadronic amplitudes for the  electromagnetic transition $I\to F$ corresponds to
the  non-local coupling of an external EM field $A^M(x,z)$  propagating in AdS with an extended fermionic mode
$\Psi_P(x,z)$, given by the l.h.s. of the equation below 
 \begin{equation} \label{FF}
 \int d^4x \, dz \,  \sqrt{g}  \,  \bar\Psi_{F, P'}(x,z)
 \,  e_M^A  \, \Gamma_A \, A^M(x,z) \Psi_{I, P}(x,z)  \\\sim
 (2 \pi)^4 \delta^4 \left( P'  \! - P \right) \epsilon_\mu  \langle \psi_F(P'), \sigma'  \vert J^\mu \vert \psi_I, \sigma \rangle,
  \end{equation} 
 where the coordinates of AdS$_5$ are the Minkowski coordinates $x^\mu$ and $z$ labeled $x^M = (x^\mu, z)$,
 with $M, N = 1, \cdots 5$, $g$ is the determinant of the metric tensor and $e^A_M$ is the vielbein with tangent indices
 $A, B = 1, \cdots, 5$.
The expression on the r.h.s.  is the EM matrix element of the quark current  $J^\mu = e_q \bar q \gamma^\mu q$ and which couples to pointlike constituents in physical space-time. As in Sec. \ref{EMFF}  one can also show that a precise mapping of the $J^+$ elements  can be carried out at fixed light-front time, providing an exact correspondence between the holographic variable $z$ and the LF impact variable $\zeta$.

 \begin{figure}[h]  
 \includegraphics[angle=0,width=7.38cm]{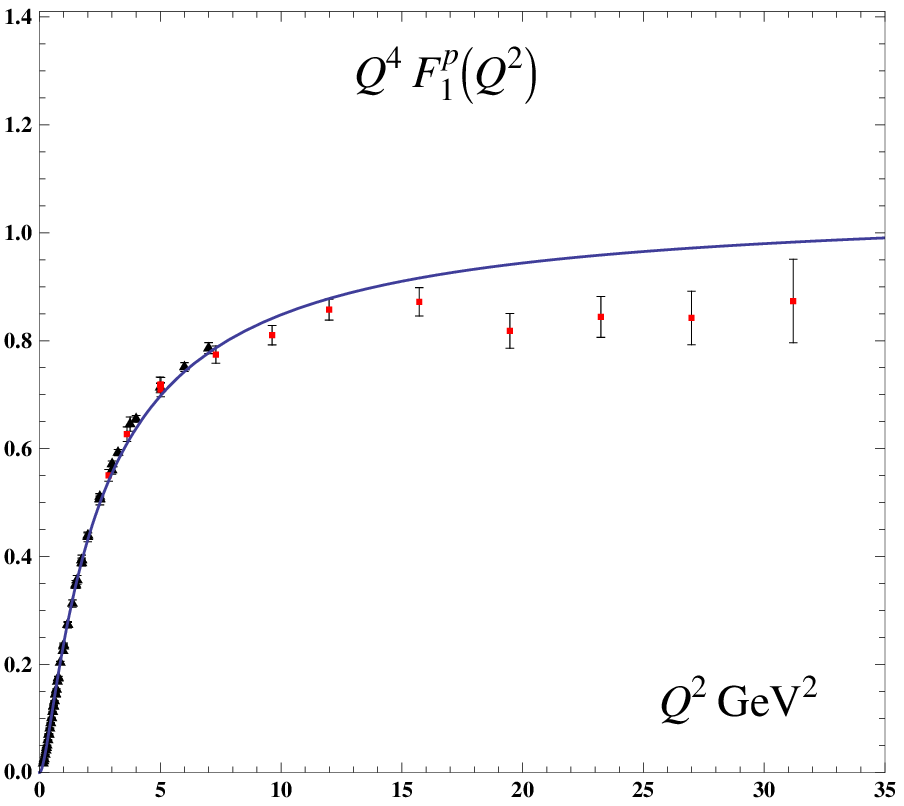}
\includegraphics[angle=0,width=7.38cm]{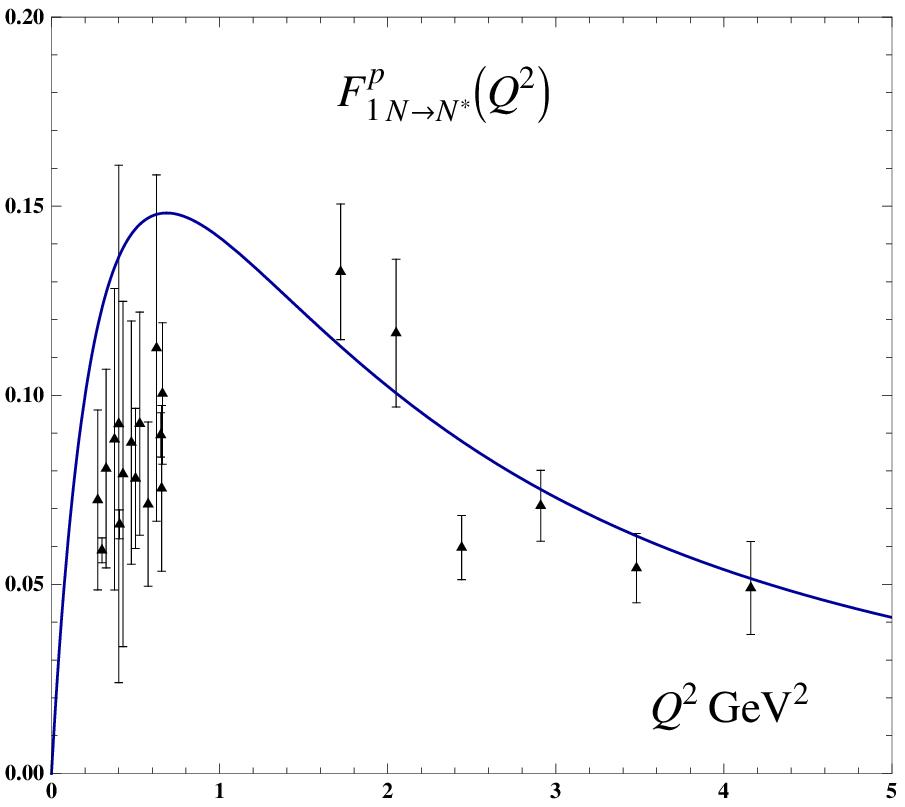}
\caption{Dirac proton form factors in light-front holographic QCD. Left: scaling of proton elastic form factor  $Q^4 F_1^p(Q^2)$. Right: proton transition form factor  ${F_1^p}_{N\to N^*}(Q^2)$ to the first radial excited state. Data compilation  from Diehl~\cite{Diehl:2005wq} (left) and JLAB~\cite{Aznauryan:2009mx}  (right).}
\label{pFFs}
\end{figure}

The proton has degenerate mass eigenstates with plus and minus components (\ref{states})
corresponding to $L^z=0$ and $L^z= + 1$ orbital components combined with spin components $S^z = + 1/2$ and $S^z = -1/2$ respectively. Likewise, we would expect that the wave equation describing the propagation of a vector meson in AdS with $J_z = + 1$ will account for three degenerate mass eigenstates with different LF orbital angular momentum components:  $L^z = 0$, $S^z = + 1$;  $L^z = +1$, $S^z = 0$ and $L^z = +2$, $S^z = -1$, which is obviously not the case in the usual formulation of AdS wave equations. To describe higher spin modes in AdS/QCD, properly incorporating the spin constituents,
the formalism  has to be extended to account for multiple component wave equations with degenerate mass eigenstates -- as for the case of the nucleon --  introducing coupled linear equations in AdS similar to the  Kemmer-Duffin-Petiau equations,  a subject worth pursuing.

\subsection{Computing Nucleon Form Factors in Holographic QCD}

In order to compute the separate features of the proton an neutron form factors, one needs to incorporate the spin-flavor structure of the nucleons,  properties  which are absent in the usual models of the gauge/gravity correspondence.
This can be readily included in AdS/QCD by weighting the different Fock-state components  by the charges and spin-projections of the quark constituents; e.g., as given by the $SU(6)$  spin-flavor symmetry. 

To simplify the
discussion we shall consider the spin-non flip proton form factors for the transition $n,L \to n' L$. Using the $SU(6)$ spin-flavor symmetry we obtain the result~\cite{Brodsky:2008pg} 
\begin{equation} \label{F}
{F_1}^p_{n, l \to n', L}(Q^2)  =    R^4 \int \frac{dz}{z^4} \, \Psi_+^{n' \!, L}(z) V(Q,z)  \Psi_+^{n, L}(z),
\end{equation}
where we have factored out the plane wave dependence of the AdS fields
\begin{equation} \label{Psip}
\Psi_+(z) = \frac{\kappa^{2+L}}{R^2}  \sqrt{\frac{2 n!}{(n+L)!}} z^{7/2+L} L_n^{L+1}\left(\kappa^2 z^2\right) 
e^{-\kappa^2 z^2/2},
\end{equation}
The bulk-to-boundary propagator
is~\cite{Grigoryan:2007my}
\begin{equation} \label{V}
V(Q,z) = \kappa^2 z^2 \int_0^1 \! \frac{dx}{(1-x)^2} \, x^{\frac{Q^2}{4 \kappa^2}} 
e^{-\kappa^2 z^2 x/(1-x)} ,
\end{equation}
with $V(Q = 0, z) = V(Q, z = 0) =1$. The orthonormality of the Laguerre polynomials in (\ref{Psip}) implies that the nucleon form factor at $Q^2 = 0$ is one if $n = n'$ and zero otherwise. Using  (\ref{V}) in (\ref{F}) we find
\begin{equation} \label{protonF1}
F_1^p(Q^2) =  \frac{1}{{\Big(1 + \frac{Q^2}{M^2_\rho} \Big) }
 \Big(1 + \frac{Q^2}{M^2_{\rho'}}  \Big) },
 \end{equation}
 for the elastic proton Dirac form factor and
 \begin{equation} \label{RoperF1}
 {F_1^p}_{N \to N^*}(Q^2) = \frac{ \sqrt{2}}{3} \frac{\frac{Q^2}{M_\rho^2}}{\Big(1 + \frac{Q^2}{M^2_\rho} \Big) 
 \Big(1 + \frac{Q^2}{M^2_{\rho'}}  \Big)
       \Big(1  + \frac{Q^2}{M^2_{\rho^{''}}} \Big)},
\end{equation}
for the EM spin non-flip proton to Roper  transition form factor.~\cite{deTeramond:2011qp} The results (\ref{protonF1}) and (\ref{RoperF1}),
 compared with available data in Fig. \ref{pFFs}, correspond to the valence approximation.   The transition form factor
 (\ref{RoperF1}) is expressed in terms of the mass of the $\rho$ vector meson and its first two radial excited states, with no additional parameters.

 \section{Higher Fock Components in Light Front Holography}
 
 The LF Hamiltonian eigenvalue equation (\ref{LFH}) is a matrix in Fock space which represents an infinite number of coupled integral equations for the Fock components $\psi_n = \langle n \vert \psi \rangle$. The resulting potential in quantum field theory can be considered as an instantaneous four-point effective interaction 
in LF time, similar to the instantaneous gluon exchange in the light-cone gauge $A^+ = 0$, which leads
to $q q \to qq$, $q \bar q \to q \bar q$, $ q \to q q \bar q$ and $\bar q \to \bar q q \bar q$, thus creating Fock states with
any number of
extra quark-antiquark pairs. In this approximation there is no mixing with  the $q \bar q g$ Fock states (no dynamical gluons) from the interaction term $g_s {\overline \psi} \gamma \cdot A \psi$ in QCD. Since models based on AdS/QCD are particularly successful in the description of exclusive processes,~\cite{Brodsky:2010cq}
this may explain the dominance of quark interchange~\cite{Gunion:1972qi}
 over quark annihilation or gluon exchange contributions in large angle elastic scattering.~\cite{Baller:1988tj}

 To show the relevance of higher Fock states we discuss  a simple semi-phenomenological model of the elastic form factor of the pion where we include the first two components in a Fock expansion of the pion wave function
$\vert \pi \rangle  = \psi_{q \bar q /\pi} \vert q \bar q  \rangle_{\tau=2}
+  \psi_{q \bar q q \bar q} \vert q \bar q  q \bar q  \rangle_{\tau=4} + \cdots$ ,
where the $J^{PC} = 0^{- +}$ twist-two and twist-4 states $\vert q \bar q \rangle$  and  $\vert q \bar q q \bar q  \rangle$ are created by the interpolating operators
$\bar q \gamma^+ \gamma_5  q$ and $ \bar q \gamma^+ \gamma_5  q  \bar q q$ respectively. 

 \begin{figure}[h]
\begin{center} 
\includegraphics[width=6.45cm]{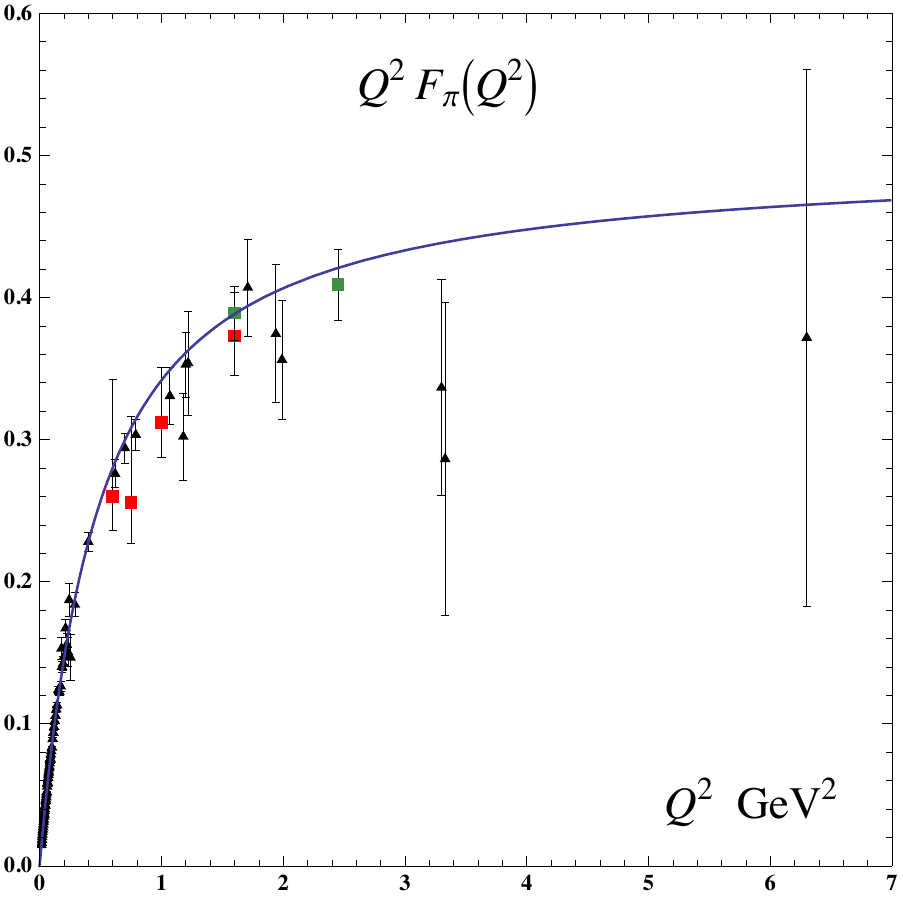} \hspace{8pt}
\includegraphics[width=7.10cm]{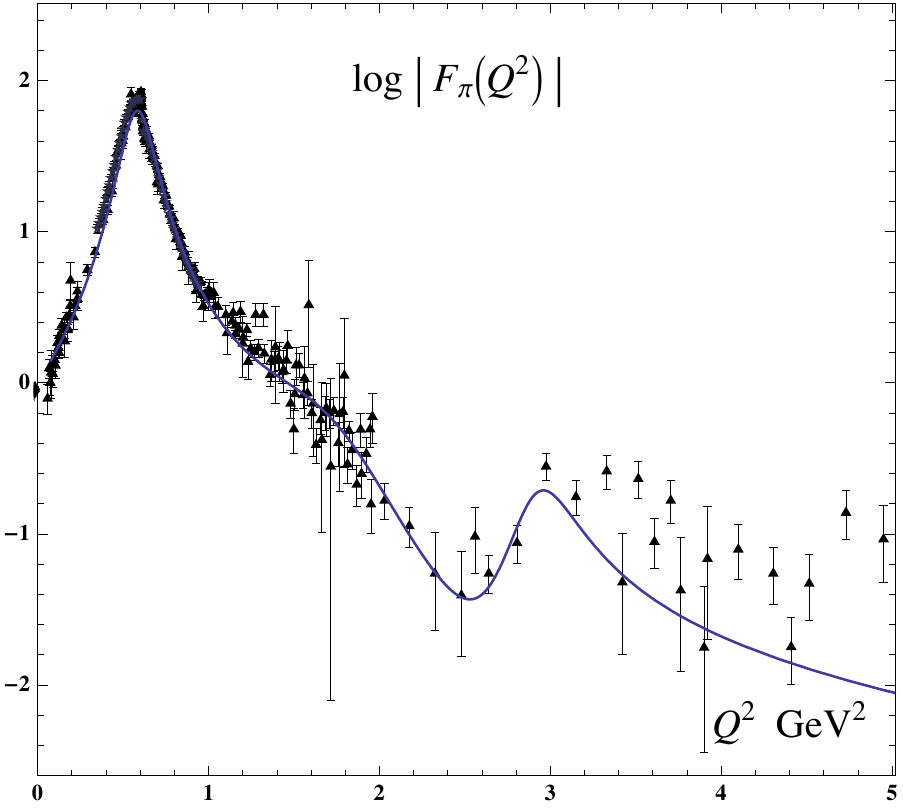}
\caption{Structure of the space- and time-like pion form factor in light-front holography for a truncation of the pion wave function up to twist four.
Triangles are the data compilation  from Baldini  {\it et al.},~\cite{Baldini:1998qn} red squares are JLAB 1~\cite{Tadevosyan:2007yd} and green squares are JLAB 2.~\cite{Horn:2006tm}}
\label{pionFFhfs}
\end{center}
\end{figure}

It is  apparent from (\ref{Ftau}) that the higher-twist components in the Fock expansion are relevant for the computation of hadronic form factors, particularly for the time-like region which is particularly sensitive to the detailed structure of the amplitudes.~\cite{deTeramond:2010ez}   Since the charge form factor is a diagonal operator, the final expression for the form factor corresponding to the truncation up to twist four is the sum of two terms, a monopole and a three-pole term.
In the strongly coupled semiclassical gauge/gravity limit hadrons have zero widths and are stable. One can nonetheless modify the formula (\ref{Ftau}) to introduce a finite width:
$q^2 \to q^2 + \sqrt 2 i M \Gamma$.  We choose the values $\Gamma_\rho =  140$ MeV,   $\Gamma_{\rho'} =  360$ MeV and  $\Gamma_{\rho''} =  120$ MeV.  The results for the pion form factor with higher Fock states (twist two and four) are shown in Fig. \ref{pionFFhfs}. The results correspond to $P_{q \bar q q \bar q}$ = 13 \%, the admixture of the
$\vert q \bar q q \bar q  \rangle$ state. The value of $P_{q \bar q q \bar q}$ (and the widths) are input in the model. The value of $\kappa$ is determined from the $\rho$ mass and the masses of the radial excitations follow from 
$M^2 \to 4 \kappa^2(n + 1/2)$. The time-like structure of the pion form factor displays a rich pole structure with constructive and destructive interferences.

\section{Novel Perspectives on QCD from Light-Front Dynamics and Light-Front Holography}
\label{NovelQCD}

In this section we summarize a number of  topics  where new, and in some cases surprising, perspectives for QCD physics have emerged from the light-front formalism and light-front holography.

\begin{enumerate}

\item  It is natural to assume that the nuclear modifications to the structure functions measured in deep inelastic lepton-nucleus and neutrino-nucleus interactions are identical;  however, the Gribov-Glauber theory predicts that the antishadowing of nuclear structure functions is not  universal, but depends on the quantum numbers of each struck quark and antiquark.~\cite{Brodsky:2004qa}  This observation can explain the recent analysis of Schienbein {\it et al.},~\cite{Schienbein:2008ay} which shows that the NuTeV measurements of nuclear structure functions obtained from neutrino  charged current reactions differ significantly from the distributions measured in deep inelastic electron and muon scattering.

\item The effects of final-state interactions of the scattered quark  in deep inelastic scattering  have been traditionally assumed to be power-law suppressed.  In fact,  the final-state gluonic interactions of the scattered quark lead to a  $T$-odd non-zero spin correlation of the plane of the lepton-quark scattering plane with the polarization of the target proton.~\cite{Brodsky:2002cx}  This  leading-twist ``Sivers effect''  is non-universal since QCD predicts an opposite-sign correlation~\cite{Collins:2002kn,Brodsky:2002rv} in Drell-Yan reactions, due to the initial-state interactions of the annihilating antiquark. 
The final-state interactions of the struck quark with the spectators~\cite{Brodsky:2002ue}  also lead to diffractive events in deep inelastic scattering (DDIS) at leading twist,  such as $\ell p \to \ell^\prime p^\prime X ,$ where the proton remains intact and isolated in rapidity;    in fact, approximately 10 \% of the deep inelastic lepton-proton scattering events observed at HERA are
diffractive.~\cite{Adloff:1997sc, Breitweg:1998gc} The presence of a rapidity gap
between the target and diffractive system requires that the target
remnant emerges in a color-singlet state; this is made possible in
any gauge by the soft rescattering incorporated in the Wilson line or by augmented light-front wavefunctions.  

\item It is usually assumed --  following the intuition of the parton model -- that the structure functions  measured in deep inelastic scattering can be computed in the Bjorken-scaling leading-twist limit from the absolute square of the light-front wavefunctions, summed over all Fock states.  In fact,  dynamical effects, such as the Sivers spin correlation and diffractive deep inelastic lepton scattering due to final-state gluon interactions,  contribute to the experimentally observed DIS cross sections. 
Diffractive events also lead to the interference of two-step and one-step processes in nuclei which in turn, via the Gribov-Glauber theory, lead to the shadowing and the antishadowing of the deep inelastic nuclear structure functions;~\cite{Brodsky:2004qa}  such phenomena are not included in the light-front wavefunctions of the nuclear eigenstate.
This leads to an important  distinction between ``dynamical''  vs. ``static''  (wavefunction-specific) structure functions.~\cite{Brodsky:2009dv}

\item
As  noted by Collins and Qiu,~\cite{Collins:2007nk} the traditional factorization formalism of perturbative QCD  fails in detail for many hard inclusive reactions because of initial- and final-state interactions.  For example, if both the
quark and antiquark in the Drell-Yan subprocess
$q \bar q \to  \mu^+ \mu^-$ interact with the spectators of the
other  hadron, then one predicts a $\cos 2\phi \sin^2 \theta$ planar correlation in unpolarized Drell-Yan
reactions.~\cite{Boer:2002ju}  This ``double Boer-Mulders effect" can account for the large $\cos 2 \phi$ correlation and the corresponding violation~\cite{Boer:2002ju, Boer:1999mm} of the Lam Tung relation for Drell-Yan processes observed by the NA10 collaboration.   
An important signal for factorization breakdown at the LHC  will be the observation of a $\cos 2 \phi$ planar correlation in dijet production.

\item	  It is conventional to assume that the charm and bottom quarks in the proton structure functions  only arise from gluon splitting $g \to Q \bar Q.$  In fact, the proton light-front wavefunction contains {\it ab initio } intrinsic heavy quark Fock state components such as $\vert uud c \bar c\rangle$.~\cite{Brodsky:1980pb,Brodsky:1984nx,Harris:1995jx,Franz:2000ee}   The intrinsic heavy quarks carry most of the proton's momentum since this minimizes the off-shellness of the state. The heavy quark pair $Q \bar Q$ in the intrinsic Fock state  is primarily a color-octet,  and the ratio of intrinsic charm to intrinsic bottom scales scales as $m_c^2/m_b^2 \simeq 1/10,$ as can easily be seen from the operator product expansion in non-Abelian QCD.   Intrinsic charm and bottom explain the origin of high $x_F$ open-charm and open-bottom hadron production, as well as the single and double $J/\psi$ hadroproduction cross sections observed at high $x_F$.   The factorization-breaking nuclear $A^\alpha(x_F)$ dependence  of hadronic $J/\psi$ production cross sections is also explained.  A novel mechanism for inclusive and diffractive
Higgs production $pp \to p p H  $, in which the Higgs boson carries a significant fraction of the projectile proton momentum, is discussed in Ref.~\cite{Brodsky:2006wb}.  The production
mechanism is based on the subprocess $(Q \bar Q) g \to H $ where the $Q \bar Q$ in the $\vert uud Q \bar Q \rangle$ intrinsic heavy quark Fock state of the colliding proton has approximately
$80\%$ of the projectile proton's momentum.

\item It is normally assumed that high transverse momentum hadrons in inclusive high energy hadronic collisions,  such as $ p p \to H X$,  can only arise  from jet fragmentation.
A  fundamental test of leading-twist QCD predictions in high transverse momentum hadronic reactions is the measurement of the power-law
fall-off of the inclusive cross section~\cite{Sivers:1975dg}
${E d \sigma/d^3p}(A B \to C X) ={ F(\theta_{cm}, x_T)/ p_T^{n_{eff}} } $ at fixed $x_T = 2 p_T/\sqrt s$
and fixed $\theta_{CM},$ where $n_{eff} \sim 4 + \delta$. Here $\delta  =  {\cal O}(1)$ is the correction to the conformal prediction arising
from the QCD running coupling and the DGLAP evolution of the input parton distribution and fragmentation functions.~\cite{Brodsky:2005fz,Arleo:2009ch,Arleo:2010yg}
The usual expectation is that leading-twist subprocesses will dominate measurements of high $p_T$ hadron production at RHIC and Tevatron energies. In fact, the  data for isolated photon production $ p p \to \gamma_{\rm direct} X,$ as well as jet production, agrees well with the  leading-twist scaling prediction $n_{eff}  \simeq 4.5$.~\cite{Arleo:2009ch}
However,   measurements  of  $n_{eff} $ for hadron production  are not consistent with the leading twist predictions.
Striking
deviations from the leading-twist predictions were also observed at lower energy at the ISR and  Fermilab fixed-target experiments.~\cite{Sivers:1975dg,Cronin:1973fd,Antreasyan:1978cw}
In fact, a significant fraction of high $p^H_\perp$ isolated hadrons can emerge
directly from hard higher-twist subprocess~\cite{Arleo:2009ch,Arleo:2010yg} even at the LHC.  The direct production of hadrons can explain~\cite{Brodsky:2008qp} the remarkable ``baryon anomaly" observed at RHIC:  the ratio of baryons to mesons at high $p^H_\perp$,  as well as the power-law fall-off $1/ p_\perp^n$ at fixed $x_\perp = 2 p_\perp/\sqrt s, $ both  increase with centrality,~\cite{Adler:2003kg} opposite to the usual expectation that protons should suffer more energy loss in the nuclear medium than mesons.
The high values $n_{eff}$ with $x_T$ seen in the data  indicate the presence of an array of higher-twist processes, including subprocesses where the hadron enters directly, rather than through jet fragmentation.~\cite{Blankenbecler:1975ct}

\item	It is often stated that the renormalization scale of the QCD running coupling $\alpha_s(\mu^2_R) $  cannot be fixed, and thus it has to be chosen in an {\it ad hoc} fashion.  In fact, as in QED, the scale can be fixed unambiguously by shifting $\mu_R$  so that all terms associated with the QCD $\beta$ function vanish.  In general, each set of skeleton diagrams has its respective scale. The result series is equivalent to the perturbative expansion of an equivalent conformal theory;  it is thus scheme independent and independent of the choice of the initial renormalization scale ${\mu_R}_0$, thus satisfying Callan-Symanzik invariance.    This is the ``principle of maximal conformality"~\cite{Brodsky:2011ig} - the principle which underlies the BLM scale setting method.
Unlike heuristic scale-setting procedures,  the BLM/PMC method~\cite{Brodsky:1982gc} gives results which are independent of the choice of renormalization scheme, as required by the transitivity property of the renormalization group.   The divergent renormalon terms of order $\alpha_s^n \beta^n n!$ are transferred to the physics of the running coupling.  Furthermore, one retains sensitivity to ``conformal''  effects which arise in higher orders; physical effects which are not associated with QCD  renormalization.  The BLM method also provides scale-fixed,
scheme-independent high precision connections between observables, such as the ``Generalized Crewther Relation'',~\cite{Brodsky:1995tb} as well as other ``Commensurate Scale Relations''.~\cite{Brodsky:1994eh,Brodsky:2000cr}  Clearly the elimination of the renormalization scale ambiguity would greatly improve the precision of QCD predictions and increase the sensitivity of searches for  new physics at the LHC.

\item It is usually assumed that the QCD coupling $\alpha_s(Q^2)$ diverges at $Q^2=0$;  i.e., ``infrared slavery''.  In fact, determinations from lattice gauge theory,  Bethe-Salpeter methods, effective charge measurements, gluon mass phenomena, and AdS/QCD all lead (in their respective scheme) to a finite value of the QCD coupling in the infrared.~\cite{Brodsky:2010ur}  Because of color confinement, the quark and gluon propagators vanish at long 
wavelength: $k < \Lambda_{QCD}$, and consequently the quantum loop corrections underlying the  QCD $\beta$-function  decouple in the infrared, and  the coupling  freezes to a finite value at 
$Q^2 \to 0$.~\cite{Brodsky:2007hb,Brodsky:2008be}   This observation underlies the use of conformal methods in AdS/QCD.

\item It is conventionally assumed that the vacuum of QCD contains quark $\langle 0 \vert q \bar q \vert 0 \rangle$ and gluon  $\langle 0 \vert  G^{\mu \nu} G_{\mu \nu} \vert 0 \rangle$ vacuum condensates, although the resulting vacuum energy density leads to a $10^{45}$  order-of-magnitude discrepancy with the 
measured cosmological constant.~\cite{Brodsky:2009zd}  However, a new perspective has emerged from Bethe-Salpeter and light-front analyses where the QCD condensates are identified as ``in-hadron'' condensates, rather than  vacuum entities, but consistent with the Gell Mann-Oakes-Renner  relation.~\cite{Brodsky:2010xf} The ``in-hadron''  condensates become realized as higher Fock states of the hadron when the theory is quantized at fixed light-front time $\tau = x^0 + x^3.$

\item  In nuclear physics nuclei are composites of nucleons. However, QCD provides a new perspective:~\cite{Brodsky:1976rz,Matveev:1977xt}  six quarks in the fundamental
$3_C$ representation of $SU(3)$ color can combine into five different color-singlet combinations, only one of which corresponds to a proton and
neutron.  The deuteron wavefunction is a proton-neutron bound state at large distances, but as the quark separation becomes smaller,
QCD evolution due to gluon exchange introduces four other ``hidden color'' states into the deuteron
wavefunction.~\cite{Brodsky:1983vf} The normalization of the deuteron form factor observed at large $Q^2$,~\cite{Arnold:1975dd} as well as the
presence of two mass scales in the scaling behavior of the reduced deuteron form factor,~\cite{Brodsky:1976rz} suggest sizable hidden-color
Fock state contributions  in the deuteron
wavefunction.~\cite{Farrar:1991qi}
The hidden-color states of the deuteron can be materialized at the hadron level as   $\Delta^{++}(uuu), \, \Delta^{-}(ddd)$ and other novel quantum
fluctuations of the deuteron. These dual hadronic components become important as one probes the deuteron at short distances, such
as in exclusive reactions at large momentum transfer.  For example, the ratio  ${{d \sigma/ dt}(\gamma d \to \Delta^{++}
\Delta^{-})/{d\sigma/dt}(\gamma d\to n p) }$ is predicted to increase to  a fixed ratio $2:5$ with increasing transverse momentum $p_T.$
Similarly, the Coulomb dissociation of the deuteron into various exclusive channels $e d \to e^\prime + p n, p p \pi^-, \Delta \, \Delta, \cdots$
will have a changing composition as the final-state hadrons are probed at high transverse momentum, reflecting the onset of hidden-color
degrees of freedom.

\item It is usually assumed that the imaginary part of the deeply virtual Compton scattering amplitude is determined at leading twist by  generalized parton distributions, but that the real part has an undetermined  ``$D$-term'' subtraction. In fact, the real part is determined by the  local  two-photon interactions of the quark current in the QCD light-front Hamiltonian.~\cite{Brodsky:2008qu,Brodsky:1971zh}  This contact interaction leads to a real energy-independent contribution to the DVCS amplitude  which is independent of the photon virtuality at fixed  $t$.  The interference of the timelike DVCS amplitude with the Bethe-Heitler amplitude leads to a charge asymmetry in $\gamma p \to \ell^+ \ell^- p$.~\cite{Brodsky:1971zh,Brodsky:1973hm,Brodsky:1972vv}   Such measurements can verify that quarks carry the fundamental electromagnetic current within hadrons.

\end{enumerate}

\section {Vacuum Effects and Light-Front Quantization\label{VacuumEffects}}

The LF vacuum is remarkably simple in light-front quantization because of the restriction $k^+ \ge 0.$   For example in QED,  vacuum graphs such as $e^+ e^- \gamma $  associated with the zero-point energy do not arise. In the Higgs theory, the usual Higgs vacuum expectation value is replaced with a $k^+=0$ zero mode;~\cite{Srivastava:2002mw} however, the resulting phenomenology is identical to the standard analysis.

Hadronic condensates play an important role in quantum chromodynamics (QCD).
Conventionally, these condensates are considered to be properties
of the QCD vacuum and hence to be constant throughout space-time.
Recently a new perspective on the nature of QCD
condensates $\langle \bar q q \rangle$ and $\langle
G_{\mu\nu}G^{\mu\nu}\rangle$, particularly where they have spatial and temporal
support,
has been presented.~\cite{Brodsky:2008be,Brodsky:2008xu,Brodsky:2009zd}
Their spatial support is restricted to the interior
of hadrons, since these condensates arise due to the interactions of quarks and
gluons which are confined within hadrons. For example, consider a meson consisting of a light quark $q$ bound to a heavy
antiquark, such as a $B$ meson.  One can analyze the propagation of the light
$q$ in the background field of the heavy $\bar b$ quark.  Solving the
Dyson-Schwinger equation for the light quark one obtains a nonzero dynamical
mass and, via the connection mentioned above, hence a nonzero value of the
condensate $\langle \bar q q \rangle$.  But this is not a true vacuum
expectation value; instead, it is the matrix element of the operator $\bar q q$
in the background field of the $\bar b$ quark.  The change in the (dynamical)
mass of the light quark in this bound state is somewhat reminiscent of the
energy shift of an electron in the Lamb shift, in that both are consequences of
the fermion being in a bound state rather than propagating freely.
Similarly, it is important to use the equations of motion for confined quarks
and gluon fields when analyzing current correlators in QCD, not free
propagators, as has often been done in traditional analyses of operator
products.  Since after a $q \bar q$ pair is created, the distance between the
quark and antiquark cannot get arbitrarily great, one cannot create a quark
condensate which has uniform extent throughout the universe.
As a result, it is argued in Refs. ~\cite{Brodsky:2008be,Brodsky:2008xu,Brodsky:2009zd}    that the 45 orders of magnitude conflict of QCD with the observed value of the cosmological condensate is removed.
A new perspective on the nature of quark and gluon condensates in
quantum chromodynamics is thus obtained:~\cite{Brodsky:2008be,Brodsky:2008xu,Brodsky:2009zd}  the spatial support of QCD condensates
is restricted to the interior of hadrons, since they arise due to the
interactions of confined quarks and gluons.  In the LF theory, the condensate physics is replaced by the dynamics of higher non-valence Fock states as shown by Casher and Susskind.~\cite{Casher:1974xd}  In particular, chiral symmetry is broken in a limited domain of size $1/ m_\pi$,  in analogy to the limited physical extent of superconductor phases.  

This novel description  of chiral symmetry breaking  in terms of ``in-hadron condensates"  has also been observed in Bethe-Salpeter studies~\cite{Maris:1997hd,Maris:1997tm}.
The usual argument for a quark vacuum condensate is the Gell-Mann--Oakes--Renner formula:
\begin{equation}
m^2_\pi = -2 m_q {\langle0| \bar q q |0\rangle\over f^2_\pi}.
\end{equation}
However, in the Bethe-Salpeter formalism, where the pion is a $q \bar q$ bound-state, the GMOR relation is replaced by
\begin{equation}
m^2_\pi = - 2 m_q {\langle 0| \bar q \gamma_5  q |\pi \rangle\over f_\pi},
\end{equation}
where $\rho_\pi \equiv - \langle0| \bar q \gamma_5  q |\pi\rangle$  represents a pion decay constant via an an elementary pseudoscalar current. The result is independent of the renormalization scale. In the light-front formalism, this matrix element derives from the $|q \bar q \rangle$ Fock state of the pion with parallel spin-projections $S^z = \pm 1$ and $L^z= \mp 1$, which couples by quark spin-flip to the usual $|q \bar q\rangle$ $S^z=0, L^z=0$ Fock state via the running quark mass. 
This new perspective explains the
results of studies~\cite{Ioffe:2002be,Davier:2007ym,Davier:2008sk} which find no significant signal for the vacuum gluon
condensate.

AdS/QCD also provides  a description of chiral symmetry breaking by
using the propagation of a scalar field $X(z)$
to represent the dynamical running quark mass.
In the hard wall model the solution has the form~\cite{Erlich:2005qh,DaRold:2005zs} $X(z) = a_1 z+ a_2 z^3$, where $a_1$ is
proportional to the current-quark mass. The coefficient $a_2$ scales as
$\Lambda^3_{QCD}$ and is the analog of $\langle \bar q q \rangle$; however,
since the quark is a color nonsinglet, the propagation of $X(z),$ and thus the
domain of the quark condensate, is limited to the region of color confinement.
Furthermore the effect of the $a_2$ term
varies within the hadron, as characteristic of an in-hadron condensate.
The AdS/QCD picture of condensates with spatial support restricted to hadrons
is also in general agreement with results from chiral bag
models,~\cite{Chodos:1975ix,Brown:1979ui,Hosaka:1996ee}
which modify the original MIT bag by coupling a pion field to the surface of
the bag in a chirally invariant manner.

\section{Conclusions}

Light-front holography provides a simple and successful framework for describing the hadronic spectrum,
the observed multiplicities and degeneracies. It also provides new analytical tools  for computing hadronic transition amplitudes, incorporating the scaling behavior and the transition to the confinement region.  The framework has a simple analytical structure and can be applied to study dynamical properties in Minkowski space-time which are not amenable to Euclidean  lattice gauge theory computations.  In spite of its present limitations,  the AdS/QCD approach, together with light-front holography, provides  important physical  insights into the non-perturbative regime of QCD and its transition to the perturbative domain where quark and gluons are the relevant degrees of freedom.

Nonzero quark masses are naturally incorporated into the AdS/LF predictions by including them explicitly in the LF kinetic energy  $\sum_i ( {\mbf{k}^2_{\perp i} + m_i^2})/{x_i}$~\cite{Brodsky:2008pg, Branz:2010ub}. Given the nonpertubative LFWFs one can predict many interesting phenomenological quantities such as heavy quark decays, generalized parton distributions and parton structure functions.  The light front AdS/QCD model is semiclassical, and thus it only predicts the lowest valence Fock state structure of the hadron LFWF. One can systematically improve the holographic approximation by
diagonalizing the QCD light front Hamiltonian on the holographic AdS/QCD basis,~\cite{Vary:2009gt} or by using the Lippmann-Schwinger equations.
The action of the non-diagonal terms
in the QCD interaction Hamiltonian also generates the form of the higher
Fock state structure of hadronic LFWFs.  In
contrast with the original AdS/CFT correspondence, the large $N_C$
limit is not required to connect light-front QCD to
an effective dual gravity approximation.

We have also reviewed some novel features  of QCD,  including
the consequences of confinement for quark and gluon condensates.
The distinction
between static structure functions, such as the probability
distributions  computed from the square of the light-front
wavefunctions, versus dynamical structure functions which include the
effects of rescattering,
has also been emphasized.
We have also discussed the relevance of the light-front Hamiltonian formulation of QCD to describe the
coalescence of quark and gluons into hadrons.

\begin{acknowledgements}
Presented by SJB and GdT at LIGHTCONE 2011, 23 - 27 May, 2011, Dallas.
We are grateful to ILCAC, Southern Methodist University, and especially Simon Dalley for organizing LC2011.
This research was supported by the Department of Energy  contract DE--AC02--76SF00515.  
SLAC-PUB-14559.
\end{acknowledgements}

\end{document}